\documentclass{llncs}

\usepackage{times}
\usepackage{latexsym}
\usepackage{pst-node}
\usepackage{amssymb,amsmath} 
\usepackage{mathrsfs}
\usepackage{color}
\usepackage{epsf}
\usepackage{pstricks,pst-plot,pst-text,pst-tree,pst-eps,pst-fill,pst-node,pst-math,pstricks-add}
\usepackage{graphicx}

\newcommand{\refe}[1] {(\ref{#1})}

\newcommand{\axiome}[1] {{\rm{\textbf{{#1}}}}}

\def\reel { \mathbb{R} }

\newcommand{\idfont}[2] {
                        \ifnum #2 = 8  \scriptsize   \fi
                        \ifnum #2 = 10 \small        \fi
                        \ifnum #2 = 11 \footnotesize \fi
                        \ifnum #2 = 12 \normalsize   \fi
                        \ifnum #2 = 14 \large        \fi
                        \ifnum #2 = 17 \Large        \fi
                        \ifnum #2 = 18 \LARGE        \fi
                        \ifnum #2 = 20 \Huge         \fi}
\definecolor{darkgreen}{rgb}{0,0.9,0}
\definecolor{darkred}{rgb}{0.9,0,0}

\begin{document}
                    
\title{Axiomatization of an importance index for $k$-ary games}

\author{Mustapha Ridaoui\inst{1}, Michel Grabisch\inst{1}, Christophe Labreuche\inst{2}}

\institute{
$^{1}$ Paris School of Economics, Universit\'e Paris I - Panth\'eon-Sorbonne, Paris, France  \\
{\tt \{mustapha.ridaoui,michel.grabisch\}@univ-paris1.fr}\\
$^{2}$ Thales Research \& Technology, Palaiseau, France\\
{\tt christophe.labreuche@thalesgroup.com}
}

\maketitle
%%%%%%%%%%%%%%%%%%%%%%%%%%%%%%%%%%%%%%%%%%%%%%%%%%%%%%%%%%%%%%%%%%%%%%%%%%%%%%%%%%%%%%%%%%
%%%%%%%%%%%%%%%%%%%%%%%%%%%%%%%%%%%%%%%%%%%%%%%%%%%%%%%%%%%%%%%%%%%%%%%%%%%%%%%%%%%%%%%%%%
%%%%%%%%%%%%%%%%%%%%%%%%%%%%%%%%%%%%%%%%%%%%%%%%%%%%%%%%%%%%%%%%%%%%%%%%%%%%%%%%%%%%%%%%%%
%%%%%%%%%%%%%%%%%%%%%%%%%%%%%%%%%%%%%%%%%%%%%%%%%%%%%%%%%%%%%%%%%%%%%%%%%%%%%%%%%%%%%%%%%%
%%%%%%%%%%%%%%%%%%%%%%%%%%%%%%%%%%%%%%%%%%%%%%%%%%%%%%%%%%%%%%%%%%%%%%%%%%%%%%%%%%%%%%%%%%
\begin{abstract}
We consider MultiCriteria Decision Analysis models which are defined over
discrete attributes, taking a finite number of values. We do not assume that the
model is monotonically increasing with respect to the attributes values. Our aim
is to define an importance index for such general models, considering that they
are equivalent to $k$-ary games (multichoice games). We show that classical solutions like
the Shapley value are not suitable for such models, essentially because of the
efficiency axiom which does not make sense in this context. We propose an
importance index which is a kind of average variation of the model along the
attributes. We give an axiomatic characterization of it.
\end{abstract}

%%%%%%%%%%%%%%%%%%%%%%%%%%%%%%%%%%%%%%%%%%%%%%%%%%%%%%%%%%%%%%%%%%%%%%%%%%%%%%%%%%%%%%%%%%
%%%%%%%%%%%%%%%%%%%%%%%%%%%%%%%%%%%%%%%%%%%%%%%%%%%%%%%%%%%%%%%%%%%%%%%%%%%%%%%%%%%%%%%%%%
%%%%%%%%%%%%%%%%%%%%%%%%%%%%%%%%%%%%%%%%%%%%%%%%%%%%%%%%%%%%%%%%%%%%%%%%%%%%%%%%%%%%%%%%%%
%%%%%%%%%%%%%%%%%%%%%%%%%%%%%%%%%%%%%%%%%%%%%%%%%%%%%%%%%%%%%%%%%%%%%%%%%%%%%%%%%%%%%%%%%%
%%%%%%%%%%%%%%%%%%%%%%%%%%%%%%%%%%%%%%%%%%%%%%%%%%%%%%%%%%%%%%%%%%%%%%%%%%%%%%%%%%%%%%%%%%
% ####################################################################
%  INTRODUCTION
% ####################################################################

% ====================================================================
% ====================================================================
\section{Introduction}
\label{Sintro}
In MultiCriteria Decision Analysis (MCDA), a central question is to determine
the importance of attributes or criteria. Suppose the preference of a decision
maker has been represented by a numerical model. For interpretation and
explanation purpose of the model, a basic requirement is to be able to assess
the importance of each attribute. If this is easy for a number of elementary
models (essentially additive ones), it becomes more challenging with
complex models. 

For models based on the Choquet integral w.r.t. a capacity or fuzzy measure (see
a survey in \cite{grla07b}), it has been recognized since a long time ago that
the Shapley value \cite{Shapley}, a concept borrowed from game theory, is the
adequate tool to quantify the importance of attributes. 

Choquet integral-based models belong to the category of decomposable models,
that is, where utility functions are defined on each attribute, and then are
aggregated by some increasing function. In this paper, we depart from this kind
of models and focus on models where there is no such separation of utilities
among the attributes. Typically, the Generalized Additive Independence (GAI)
model proposed by Fishburn \cite{fis67,fis70} is of this type, since of the form
$U(x)=\sum_{S\in\mathcal{S}}u_S(x_S)$, where $\mathcal{S}$ is a collection of
subsets of $N$, the index set of all attributes. In this paper, however, we do
not take advantage of this peculiar form, and consider a numerical model without
particular properties, except that the underlying attributes are discrete, and
thus take a finite number of values. Note that in many applications, especially
in the AI field, this is the case, in particular for GAI models
\cite{bacgrov95,gonperndub11,brazbout07}. 

As far as we know, the question of the definition of an importance index for
such a general case remains open. As we will explain, discrete models can be
seen as $k$-ary capacities or more generally $k$-ary games \cite{GrabLab2003} (also called
multichoice games \cite{Hsiao1990}), and thus it seems natural to take as
importance index the various definitions of Shapley-like values for multichoice
games existing in the literature. There is however a major drawback  inherent to
these values: they all satisfy the efficiency axiom, that is, the sum of the
importance indices over all attributes is equal to $v(k_N)$, the value of the
game when all attributes take the highest value. If this axiom is natural in a
context of cooperative game, where the Shapley value defines a rational way to
share among the players the total benefit $v(k_N)$ of the game, it has no
justification in MCDA, especially if the model $v$ is not monotone increasing. 

The approach we propose here is inspired by the calculus of variations: we define the
importance index of an attribute as the average variation of $v$ (depicting the
satisfaction of the decision maker) when the value of attribute $i$ is increased
by one unit. We propose an axiomatic definition, where the chosen axioms are
close to those of the original Shapley value. 

Section \ref{Spre} recalls the basic concepts.
Section \ref{Simp} informally defines  what is the aim of our importance index.
The axiomatic characterization is presented in Section \ref{Saxiom}.
The new index is then interpreted (Section \ref{Sinter}).
Finally we compare our approach to related works (Section \ref{Srela}).

% ====================================================================
% ====================================================================
%%%%%%%%%%%%%%%%%%%%%%%%%%%%%%%%%%%%%%%%%%%%%%%%%%%%%%%%%%%%%%%%%%%%%%%%%%%%%%%%%%%%%%%%%%
%%%%%%%%%%%%%%%%%%%%%%%%%%%%%%%%%%%%%%%%%%%%%%%%%%%%%%%%%%%%%%%%%%%%%%%%%%%%%%%%%%%%%%%%%%
\section{Preliminaries}
\label{Spre}

Throughout the paper, $N=\lbrace 1, \ldots, n\rbrace$ is a finite set which can be thought as the set of attributes (in MCDA), players (in cooperative game theory), etc., depending on the application. 
In this paper, we will mainly focus on MCDA applications.
Cardinality of sets will be often denoted by corresponding lower case letters, e.g., $n$ for $|N|$, $s$ for $|S|$, etc.

The set of all possible values taken by attribute $i\in N$ is denoted by $L_i$.
As it is often the case in MCDA, we assume that these sets are finite, and we
represent them by integer values, i.e.,  $L_i=\lbrace 0, 1, \ldots, k^i\rbrace$.
Alternatives are thus elements of the Cartesian product $L=\times_{i\in N} L_i$
and take the form $x = (x_1, x_2, \ldots, x_n)$ with $x_i \in L_i$,
$i=1,\ldots,n$.  For $x,y\in L$, we write $x\leq y$ if $x_i\leq
y_i$ for every $i\in N$.  For $S\subseteq N$ and $x\in L$, $x_S$ is the
restriction of $x$ to $S$.  $L_{-i}$ is a shorthand for $\times_{j\neq i}L_j$. For each
$y_{-i}\in L_{-i}$, and any $\ell\in L_i$, $(y_{-i}, \ell_i)$ denotes the
combined alternative $x$ such that $x_i=\ell_i$ and $x_j=y_j, \forall j\neq
i$. The vector $0_N = (0,\ldots, 0)$ is the null alternative of $L$, and
$k_N=(k^1_1, \ldots, k^n_n )$ is the top element of $L$. $0_{-i}$ denotes the
element of $L_{-i}$ in which all coordinates are zero. We call vertex of $L$ any
element $x\in L$ such that $x_i$ is either $0$ or $k^i$, for each $i\in N$. We
denote by $\Gamma (L) = \times_{i \in N} \lbrace 0, k^i \rbrace $ the set of
vertices of $L$. For each $x \in L$, we denote by $S(x)=\lbrace i\in N \mid
x_i>0\rbrace$ the support of $x$, and by $K(x)=\lbrace i\in N \mid x_i=k^i\rbrace$
the kernel of $x$. Their cardinalities are respectively denoted by $s(x)$ and $k(x)$.

We suppose to have a numerical representation $v:L \rightarrow \reel$ of the
preference of the decision maker (DM) over the set of alternatives in $L$. For the sake
of generality, we do not make any assumption on $v$, except that $v(0_N)=0$
(this is not a restriction, as most of numerical representations are unique up
to a positive affine transformation). In particular, there is no assumption of
monotonicity, that is, we do not assume that $x_1\geq
x'_1$,\ldots, $x_n\geq x'_n$ implies $v(x_1,\ldots,x_n)\geq v(x'_1,\ldots,
x'_n)$. Example~\ref{Ex1} below illustrates that it is quite common to observe
this lack of monotonicity.
\begin{example}
 The level of comfort of humans depends on three main attributes: temperature of the air ($X_1$), humidity of the air ($X_2$)
 and velocity of the air ($X_3$). Then $v(x_1,x_2,x_3)$ measures the comfort level.
 One can readily see that $v$ is not monotone in its three arguments.
 For $x_2$ and $x_3$ fixed, $v$ is maximal for intermediate values of the
 temperature (typically around $23^\circ$C). Similarly, the value of humidity maximizing $v$ is neither too low nor too high.
 Finally, for $x_1$ relatively large, some wind is well appreciated, but not too much. 
 Hence for any $i$, and supposing the other two attributes being fixed, there
 exists an optimal value $\widehat{\ell}_i\in L_i$ such that $v$ is increasing in $x_i$ below $\widehat{\ell}_i$, and then decreasing in $x_i$ above $\widehat{\ell}_i$.
\label{Ex1}
\end{example}

Although we will not use this specific form for $v$ in the sequel, we mention as
typical example of a model not necessarily satisfying monotonicity the
Generalized Additive Independence (GAI) model, i.e., $v$ is written as
$v(x)=\sum_{S\in \mathcal{S}} v_S(x_S)$, where $\mathcal{S}$ is a collection of
subsets of $N$ \cite{fis67,fis70}.  This model has been widely used in AI
\cite{bacgrov95,gonperndub11,brazbout07}.

For convenience, we assume from now on that all attributes have the same number of elements,
i.e., $k^i=k$ for every $i\in N$ ($k\in \mathbb{N}$).  Note that if this is not
the case, $k$ is set to $\max_{i\in N} k^i$, and we duplicate some elements of
$L_i$ when $k^i<k$. A fundamental observation is that when $k=1$, $v$ is nothing
other than a \textit{pseudo-Boolean function} $v:\{0,1\}^N\rightarrow\reel$
vanishing on $0_N$, or put otherwise via the identity between sets and their
characteristic functions, a \textit{(cooperative) game} (in characteristic form)
$\mu : 2^N \rightarrow \mathbb{R}$, with $\mu(\varnothing) = 0$.  A game $v$ is
\textit{monotone} if $v(A)\leq v(B)$ whenever $A\subseteq B$. A monotonic game
is called a \textit{capacity} \cite{Choquet} or \textit{fuzzy measure}
\cite{Sugeno1}. For the general case $k\geq 1$, $v:L\rightarrow \reel$ is
called a \textit{multichoice game} or \textit{$k$-ary game} \cite{Hsiao1990}, and the
numbers $0,1,\ldots,k$ in $L_i$ are seen as the level of activity of the
players. By analogy with the classical case $k=1$, a \textit{$k$-ary capacity}
is a monotone $k$-ary game, i.e., satisfying $v(x)\leq v(y)$ whenever $x\leq y$,
for each $x, y \in L$ \cite{GrabLab2003}. Hence, a $k$-ary capacity represents a
preference on $L$ which is increasing with the value of the attributes.  We
denote by $\mathcal{G}(L)$ the set of $k$-ary games defined on $L$.
    
The M\"obius transform of a $k$-ary game $v$  is a mapping $m^v:L\rightarrow \mathbb{R}$ which is the unique solution of the linear system (cf. \cite{Rota})
\begin{equation}
\label{eq:game}
v(x)=\sum_{y\leq x} m^{v}(y), x\in L.
\end{equation}
Its solution is given by
%\begin{equation}
%\label{eq:mob}
$m^{v}(x)=\displaystyle\sum_{\substack{y\leq x\\x_i - y_i\leq 1,\:\forall i\in N}}(-1)^{\sum_{i\in N} (x_i - y_i)}v(y) , x\in L$.\\
\\
%\end{equation}
By analogy with classical games, a unanimity game for $k$-ary game denoted $u_x$, for each $x\in L$ with $x\neq 0_N$ is defined by
\begin{equation*}
u_x(y)=\left\lbrace
\begin{array}{ll}
1, & \mbox{if $y \geq x$}\\
0, & \mbox{otherwise}
\end{array}
\right.
\end{equation*}
Hence, (\ref{eq:game}) can be rewritten as
\begin{equation}
\label{eq:gameU}
v = \sum_{\substack{x\in L\\x\neq 0_N}}m^v(x)u_x
\end{equation}
Note that the set of unanimity games forms a basis of the vector space of $k$-ary games.
 One advantage of unanimity games is that they are monotone. Hence this basis is relevant for $k$-ary capacities.
In order to obtain a basis of $k$-ary games not necessarily made of monotone functions,
we define for each $x\in L$ such that $x\neq 0_N$, the game $\delta_x$ by 
\begin{equation*}
\delta_x(y)=\left\lbrace
\begin{array}{ll}
1, & \mbox{if $y = x$}\\
0, & \mbox{otherwise}
\end{array}
\right.
\end{equation*}
It is obvious that any $k$-ary game $v$ can be written as
\begin{equation}
v = \sum_{\substack{x\in L\\x\neq 0_N}} v(x)\delta_x.
\end{equation}

% ====================================================================
% ====================================================================
\section{Definition of an importance index: what do we aim at doing?}
\label{Simp}

When dealing with numerical representations of preference in MCDA, one of the
primary concerns is to give an interpretation of the model in terms of
importance of the attributes. When $v$ is a capacity or a game ($k=1$), or with
continuous models extending capacities and games like the Choquet integral, the
standard solution is to take the Shapley value, introduced by Shapley in the
context of cooperative games \cite{Shapley}. A \textit{value} is a function
$\phi :  \mathcal{G}(2^N)\rightarrow \mathbb{R}^N$ that assigns to every
game $\mu$ a payoff vector $\phi(\mu)$. It is interpreted in the MCDA context as the
vector of importance of the attributes. The value introduced by Shapley is one
of the most popular, and is defined by: 
\begin{equation}
\phi_i^{Sh}(\mu) = \sum_{S\subseteq N\setminus i} \frac{(n-s-1)!s!}{n!} \big(\mu(S\cup i)-\mu(S)\big),  \forall i\in N.
\end{equation}
A standard property shared by many values in the literature is
\textit{efficiency}: $\sum_{i\in N}\phi_i(\mu)=\mu(N)$. This property is very
natural in game theory, as $\mu(N)$ is the total benefit obtained from the
cooperation of all players in $N$, and by efficiency the payoff vector $\phi(v)$
represents a sharing of this total benefit.

If the Shapley value has been widely used in MCDA with great success (see, e.g.,
\cite{grla07b}), it must be stressed that it was only in the case of
monotonically increasing
models, i.e., based on a capacity $\mu$. In such cases, $\mu(N)$ is set to 1,
the value of the best possible alternative, and the importance index of an
attribute could be seen as a kind of contribution of that attribute to the best
possible alternative. However if the model is not monotone increasing, such an
interpretation fails. Hence, we are facing here a double difficulty: to propose
a ``value'' both valid for $k\geq 1$ and for nonincreasing
models. Section~\ref{Srela} presents several definitions found in the literature
of a Shapley-like value for $k$-ary games. However, all of them satisfy the
efficiency axiom. 

 We wish to capture in our importance index the impact of each attribute on the
 overall utility. Let us consider for illustration the game $\delta_y$ with
 $k=2$, $n=3$ and $y=(2,1,1)$.  Attribute $1$ is non-decreasing and has a
 positive impact on the overall utility.  Attributes $2$ and $3$ have neither a
 positive nor a negative impact on the overall utility, since $\delta_y(x)$ is
 non-decreasing (resp., non-increasing) when $x_2$ or $x_3$ goes from $0$ to $1$
 (resp., from $1$ to $2$).  Hence, denoting by $\phi(\delta_y)$ our importance
 index for that game, one shall have $\phi_1(\delta_y)>0$,
 $\phi_2(\delta_y)=\phi_3(\delta_y)=0$, so that the sum $\sum_{i\in N} \phi_i(v)
 >0$ cannot be equal to $v(2,2,2)-v(0,0,0)=0$, and hence $\phi$ does not satisfy
 efficiency. Rather, the index $\phi_i(v)$ shall measure the impact of attribute
 $i$ on $v$, as the total variation on $v$ if we increase
 the value of attribute $i$ of one unit (going from value $x_i$ to $x_i+1$),
 when $x$ is varying over the domain.

% ======================================================================================================================
% ======================================================================================================================
\section{Axiomatization}
\label{Saxiom}

We define in this section an importance index according to the ideas explained
above, by using an axiomatic description. Our axioms are relatively close to the
ones used by Shapley when characterizing his value in \cite{Shapley}: linearity,
null player, symmetry and efficiency. Our approach will follow Weber
\cite{Weber}, who introduces the axioms one by one and at each step gives a
characterization. Throughout this section, we consider a value as a mapping
$\phi:\mathcal{G}(L)\rightarrow \reel$. 

\medskip

The linearity axiom means that if we have the preferences $v$ and $w$ of two
DMs, and the resulting preference is a linear combination of them (yielding
$v=\alpha \: v + \beta \: w$), then it is equivalent to apply $\phi$ before or
after the linear combination.  Axiom \axiome{L} is also very helpful in the
view of the GAI decomposition.  
\begin{quote}
\textbf{Linearity axiom (L) }: $\phi$ is linear on $\mathcal{G}(L)$, i.e., $\forall v, w \in \mathcal{G}(L), \forall\alpha\in\mathbb{R},$ 
$$\phi_i(v+\alpha w)=\phi(v)+\alpha\phi(w).$$
\end{quote}
\begin{proposition} \label{PROP L}
Under  axiom \textbf{(L)}, for all $i\in N$, there exists constants  $ a^i_x \in\mathbb{R}$, for all $x\in L$, such that $\forall v\in\mathcal{G} (L),$   
\begin{align}
\label{L}
\phi_i(v)=\sum_{x\in L} a_x^i v(x)
\end{align}
\end{proposition}
The proof of this result and the other ones are omitted due to space limitation.

The second axiom that characterizes the Shapley value in \cite{Weber} is called
the null player axiom. It says that a player $i\in N$ who brings no contribution
(i.e., $\mu(S\cup i)=\mu(S), \forall S\subseteq N\setminus \lbrace i \rbrace$)
should receive a zero payoff. This definition can be easily extended to $k$-ary games.  
\begin{definition} A player $i\in N$ is said to be null for $v\in\mathcal{G} (L)$ if
$$v(x+1_i)=v(x), \forall x \in L, x_i<k.$$
\end{definition}
\begin{remark}
Let $i\in N$ be a null player for $v\in\mathcal{G} (L)$. we have, 
$$\forall x\in L, v(x_{-i}, x_i)=v(x_{-i}, 0_i).$$
\end{remark}
If an attribute is null w.r.t. a game $v$, then this attribute has no influence
on $v$, and hence the importance of this attribute shall be zero.
We propose the following axiom.

\begin{quote}
\textbf{Null axiom (N):} If a player $i$ is null for $v\in\mathcal{G} (L)$, then $\phi_i(v)=0$.
\end{quote}

\begin{proposition} \label{PROP LN}
Under axioms \textbf{(N)} and \textbf{(L)}, for all $i\in N$, there exists $ p^i_x \in\mathbb{R}$, for all $x\in L$ with $x_i<k$, such that $\forall v\in\mathcal{G} (L),$   
\begin{align}
\label{LN}
\phi_i(v)=\sum_{\substack{x\in L\\x_i < k}}p_x^i \big(v(x+1_i)-v(x)\big)
\end{align}
\end{proposition}
This proposition shows that $\phi_i$ is a linear combination of the added-values on $v$, going from
value $x_i$ to $x_i+1$, over all $x$.

The classical symmetry axiom says that the numbering of the attributes has no
influence on the value. It means that the computation of value should not depend on the numbering of the attributes. 

Let $\sigma$ be a permutation on $N$. For all $x\in L$, we denote $\sigma(x)_{\sigma(i)}=x_i$. For all $v\in\mathcal{G} (L)$, The game $\sigma\circ v$ is defined by $\sigma\circ v (\sigma (x))=v(x)$.

\begin{quote}
\textbf{Symmetry axiom (S):} For any permutation $\sigma$ of $N$, $\phi_{\sigma(i)}(\sigma\circ v)=\phi_i(v), \forall i\in N.$
\end{quote}

\begin{proposition} \label{PROP LNS}
Under axioms \textbf{(N)}, \textbf{(L)} and \textbf{(S)}, $\forall v\in\mathcal{G} (L), \forall i\in N$,  
$$\phi_i(v)=\sum_{\substack{x\in L\\x_i < k}}p_{x_i; n_0,\ldots, n_k} \big(v(x+1_i)-v(x)\big)$$
where $p_{x_i; n_0,\ldots, n_k} \in\mathbb{R}$, and $n_j$ is the number of components of $x_{-i}$ being equal to $j$.
\end{proposition}
This result means that the coefficients in front of the added-values on $v$, going from
value $x_i$ to $x_i+1$, do not depend on the precise value of $x$, but only on the number of terms of $x_{-i}$ taking values $0,1,\ldots,k$.

The next axiom enables an easier computation of coefficients $p_x^i$ while reducing their number.

\begin{quote}
\textbf{Invariance axiom (I):} Let us consider two games $v, w \in\mathcal{G} (L) $ such that, for all $i \in N$,

$$v(x+1_i)-v(x) = w(x)- w(x-1_i), \forall x\in L, x_i\notin \lbrace 0, k\rbrace$$
$$v(x_{-i}, 1_i) - v(x_{-i}, 0_i) = w(x_{-i}, k_i) - w(x_{-i}, k_i-1), \forall x_{-i}\in L_{-i}.$$
Then $\phi_i(v) = \phi_i(w)$.
\end{quote}

Taking two games $v$ and $w$ for which the differences $v(x+1_i)-v(x)$
(measuring the added value of improving $x$ of one unit on attribute $i$) can be
deduced from that of $w$ just by shifting of one unit, then the mean importance
of attribute $i$ shall be the same for $v$ and $w$.  In other words, what is
essential is the absolute value of the differences $v(x+1_i)-v(x)$ and not the
value $x$ at which it occurs.  
%In other words, the axiom says that when the
%contribution of attribute $i$ to a game $v$ is merely a shift of that to another
%game $w$, the values are the same for this player.

\begin{proposition} \label{PROP LNI}
Under axioms \textbf{(L)}, \textbf{(N)} and \textbf{(I)}, $\forall v\in\mathcal{G} (L), \forall i\in N$,  
$$\phi_i(v)=\sum_{x_{-i}\in L_{-i}}p_{x_{-i}}^i \big(v(x_{-i}, k_i)-v(x_{-i}, 0_i)\big)$$
\end{proposition}
Axiom \textbf{(I)} implies that we only need to look at the difference of $v$ between the extreme value $0$ and $k$. 
The evaluatinon on the intermadiate elements of $L_i$ do not count.

\begin{proposition} \label{PROP LNIS}
Under axioms \textbf{(L)}, \textbf{(N)}, \textbf{(I)} and \textbf{(S)}, $\forall v\in\mathcal{G} (L), \forall i\in N$,  
$$\phi_i(v)=\sum_{x_{-i}\in L_{-i}}p_{n(x_{-i})} \big(v(x_{-i}, k_i)-v(x_{-i}, 0_i)\big),$$
\end{proposition}
where $n(x_{-i})=(n_0,n_1,\ldots,n_k)$ with $n_j$ the number of components of
$x_{-i}$ being equal to $j$.

As explained in Section~\ref{Simp}, we do not require that $\phi$ satisfy
efficiency.  In the context of game theory, $\phi_i^{Sh}(\mu)$ is the amount of
money alloted to player $i$, so that relation $\sum_{i\in N} \phi_i^{Sh}(\mu) =
\mu(N)$ means that all players share among themselves the total worth $\mu(N)$.
We have no such interpretation in MCDA.  By contrast, we interpret $\phi_i(v)$
as an overall added value when increasing the value of attribute $i$ of one unit
-- thereby going from any point $x$ to $(x_i+1,x_{-i})$.  Hence $\sum_{i\in N}
\phi_i(v)$ can be interpreted as the overall added value when increasing
simultaneously the value of all attributes of one unit -- thereby going from any
point $x$ to $x+1=(x_1+1,\ldots,x_n+1)$.  For an arbitrary game $v$, there is a
priori no particular property for the previous sum.  We thus consider a very
special case of games following Example~\ref{Ex1}.  These games are single
peaked. The simplest version of these games is the family of games
$\delta_y$. For those games, we immediately see from Proposition~\ref{PROP LNIS}
that $\phi_i(\delta_y)=0$ for every $i$ such that $y_i\neq 0,k$, as already mentioned in Section \ref{Simp}. Based on this
remark, we should only bother on attributes which are equal to either 0 or $k$
in $y$. We have therefore three cases (recall that $s(y), k(y)$ are the
cardinalities of the support and kernel of $y$):
\begin{itemize}
\item $k(y) \not= 0$ and $s(y)=n$. Then  $y-1\in L$ because no component of $y$ is
  equal to 0, and we have $\delta_y(y)-\delta_y(y-1)=1$. Note that
  $\delta_y(x+1)-\delta_y(x)=0$ for any $x\neq y-1$ and $x,x+1\in L$. Therefore,
  by the above argument, we have $\sum_{i\in N}\phi_i(\delta_y)=1$. 
\item $k(y)=0$ and $s(y)<n$. This is the dual situation: $y+1\in L$ because no
  component is equal to $k$, and we have $\delta_y(y+1)-\delta_y(y)=-1$. Since
  $\delta_y(x+1)-\delta_y(x)=0$ for any other possible $x$, we get $\sum_{i\in
    N}\phi_i(\delta_y)=-1$.
\item $k(y) \not= 0$ and $s(y)<n$. This time there are both components equal to
  0 and to $k$ in $y$. Therefore, neither $y+1$ nor $y-1$ belong to $L$, and for
  any possible $x\in L$ s.t. $x+1\in L$, we have
  $\delta_y(x+1)-\delta_y(x)=0$. Therefore, $\sum_{i\in N}\phi_i(\delta_y)=0$.
\end{itemize}
To summarize, we shall write
\[ \sum_{i \in N} \phi_i(\delta_x) =  \left\{ \begin{array}{l}
     +1 \quad \mbox{if $k(y) \not= 0$ and $s(y)=n$}  \\ 
     -1 \quad \mbox{if $k(y)=0$ and $s(y)<n$}  \\ 
     0  \quad \mbox{else}
   \end{array}  \right.
\]
This can be written in the following compact form. 

\begin{quote}
\textbf{Efficiency axiom (E):} For all $x \in L\setminus 0$, 
$$\sum_{i \in N} \phi_i(\delta_x) = \delta_x(x_{-i}, k_i)-\delta_x(x_{-j}, 0_j)$$
where, $i=argmax\ x$ and $j=argmin\ x$
\end{quote}

Note that the previous formula takes the form of standard efficiency $\sum_{i\in N} \phi_i^{Sh}(\mu) = \mu(N) - \mu(\emptyset)$.
The final result is the following.
\begin{theorem}
Under axioms \textbf{(L)}, \textbf{(N)}, \textbf{(I)}, \textbf{(S)} and \textbf{(E)}, for all $v\in\mathcal{G} (L)$
$$\phi_i(v) = \sum_{x_{-i}\in L_{-i}} \frac{(n-s(x_{-i})-1)!k(x_{-i})!}{(n+k(x_{-i})-s(x_{-i}))!}\big(v(x_{-i}, k_i)-v(x_{-i}, 0_i)\big), \forall i\in N$$
\end{theorem}

We note that we have the following relation, for every $v\in\mathcal{G} (L)$
\[ \sum_{i \in N} \phi_i(v)=\sum_{\substack{x\in L\\x_j<k}}\big(v(x+1)-v(x)\big) .
\]
The right-hand side of this expression corresponds exactly to the interpretation provided above saying that
 $\sum_{i\in N} \phi_i(v)$ is the overall impact of going from any point $x$ to $x+1$.

% ==================================================================================
\section{Interpretation}
\label{Sinter}

We propose here an interpretation of $\phi$ in continuous spaces, that is, after
extending $v$ to the continuous domain $[0,k]^N$.  We consider thus a function
$U:[0,k]^N \rightarrow \reel$ which extends $v$: $U(x)=v(x)$ for every $x\in L$.
The importance of attribute $i$ can be defined as (see \cite[proposition 5.3.3
  page 141]{mar98}) 
% and also \cite[Definition 10.41 and Proposition 10.43 page 369]{gramamepa09})
\[ \mathit{Imp}_i(U) = \int_{[0,k]^{n-1}} \Big( U(k_i,z_{-i}) - U(0_i,z_{-i}) \Big) dz_{-i}
   =  \int_{[0,k]^n} \frac{\partial U}{\partial z_i}(z) \: dz .
\]
In this formula, the local importance of attribute $i$ for function $U$ at point
$z$ is equal to $\frac{\partial U}{\partial z_i}(z)$.  The index $
\mathit{Imp}_i(U)$ appears as the mean of relative amplitude of the range of $U$
w.r.t. attribute $i$, when the remaining variables take uniformly random values.

The most usual extension of $v$ on $[0,k]^N$ is the Choquet integral with respect to $k$-ary capacities \cite{gralab03r}.
Let us compute $ \mathit{Imp}_i$ in this case. We write
$ \mathit{Imp}_i(U) = \sum_{x\in \{0,\ldots,k-1\}^N}$ $\int_{[x,x+1]^n} \frac{\partial U}{\partial z_i}(z) \: dz $.
In $[x,x+1]^n$, $U$ is equal to $v(x)$ plus the Choquet integral $C_{\mu_x}$ w.r.t. capacity $\mu_x$ defined by 
$\mu_x(S) = v((x+1)_S,x_{-S})-v(x)$ for every $S\subseteq N$.
By \cite{mar98}, $\int_{[0,1]^n} \frac{\partial C_{\mu_x}}{\partial z_i}(z) \: dz = \phi^{Sh}_i(\mu_x)$.
Hence 
\begin{equation}
 \mathit{Imp}_i(U) = \sum_{x\in \{0,\ldots,k-1\}^N} \phi^{Sh}_i(\mu_x) .
 \label{Edecomp} 
\end{equation}
We then obtain the following result.
% It is easy to show that our index formulae $\phi_i(v)$ turns out to be equal to $\mathit{Imp}_i(U)$, when $U$ is the Choquet integral with respect to $k$-ary capacity $v$.
\begin{lemma}
  If $U$ is the Choquet integral w.r.t. $k$-ary capacity $v$, then $\mathit{Imp}_i(U) = \phi_i(v)$.
\label{L1}
\end{lemma}
Hence the counterpart of $\phi_i$ on continuous domains is the integrated local importance.

% We note that $\phi_i$ is scale invariant. Consider a function $F:[0,1]^N \rightarrow \reel$ and define the $k$-ary game by $v^k(x)=F\pp{\frac{x}{k}}$. Then $\phi_i(v^k)$ is independent of $k$. It is independent on the number of points in the grid.

\section{Related works}\label{Srela}
There have been many proposed values for multichoice games, e.g., Hsiao and
Raghavan \cite{Hsiao1990}, van den Nouweland et al. \cite{vandenNouweland1991},
Klijn et al. \cite{Klijn1999}, Peters and Zank \cite{Peters2005} and Grabisch
and Lange \cite{Grabisch2007}. We present in this section the Shapley value
defined by Hsiao and Raghavan, Peters and Zank, and Grabisch and Lange. All of
them satisfy the classical efficiency axiom and thus differ from our value.

The first extension of the Shapley value was introduced by Hsiao and Raghavan
\cite{Hsiao1990}. They defined the Shapley value using weights for all possible actions of the players, thereby extending ideas of weighted Shapley values (cf. \cite{Kalai1987}). The value proposed by Hsiao and Raghavan is based on unanimity games. They propose the following definition:
\begin{equation*}
\forall x\in L\setminus 0_N, \phi_{ij}(u_x)=\left\lbrace
\begin{array}{ll}
\frac{w_j}{\sum_{i\in N} w_{x_i}}, & \mbox{if $j = x_i$}\\
0, & \mbox{otherwise}
\end{array}
\right.
\end{equation*}
where $w_1\ldots, w_k$ are the weights of actions $1,\ldots, k$, such that $w_1<\ldots<w_k$.
Furthermore, the value is determined by $$\phi(v) = \sum_{\substack{x\in L\\x\neq 0_N}} m^v(x) \phi(u_x), \forall v\in \mathcal{G}(L),$$
where $m^v$ is the M\"obius transform of $v$.

The axiomatic of Peters and Zank \cite{Peters2005} is also based on unanimity games. They proposed the following multi-choice Shapley value,
\begin{equation*}
\phi_{ij}(v) = \sum_{\substack{x\in L\\x_i=j}}\frac{m^v(x)}{s(x)}, \forall v\in \mathcal{G}(L).
\end{equation*}

Grabisch and Lange \cite{Grabisch2007} did not use unanimity games, but took an axiomatic approach to define a Shapley value in a more general context for games over lattices. They define the Shapley value for multichoice game as follows,
$$\phi_i(v)=\sum_{x_{-i}\in\Gamma(L_{-i})}\frac{(n-k(x_{-i})-1)!k(x_{-i})!}{n!}(v(x_{-i}, k_i)-v(x_{-i}, 0_i)).$$

% ==================================================================================
\section{Conclusion and related works}

We have proposed a new importance index for $k$-ary games.  It quantifies the
impact of each attribute on the overall utility.  According to the linearity,
dummy player, symmetry and invariance properties, $\phi_i(v)$ takes the form of
the sum over $x\in \{0,\ldots,k-1\}^N$ of a value over the restriction of the
$k$-ary game on $\times_{i\in N}\{x_i,x_i+1\}$ (see \refe{Edecomp}).  In our
construction, the value at an elementary cell $\times_{i\in N}\{x_i,x_i+1\}$
corresponds to the usual Shapley value.  We will explore in future work the
possibility of the use of other values such as the Banzhaf value.  We will also
investigate other indices $\phi_i$ which measure the impact in absolute value of
attribute $i$.  In this case, $\phi_i(\delta_y)$ is not equal to zero when
$0<y_i<k$.

\bibliographystyle{abbrv}
\bibliography{References}

\end{document}